# Gamified Requirement Elicitation for a Multi-Modal Decision Support System. The Case of SYNCHROMODE.

Dimitris Tzanis [1[0000-0002-6481-5204]], Alexandros Dolianitis [1[0000-0001-8627-1968]], Viktoria Petkani[1[0009-0002-4795-5106]], Areti Kotsi[1[0000-0001-5623-0575]] and Evangelos Mitsakis [1[0000-0001-8421-9226]]

[1] Centre for Research and Technology Hellas - Hellenic Institute of Transport, 6th km Charilaou-Thermi Rd, 57001 Thermi, Thessaloniki, Greece
certh@certh.gr

**Abstract.** SYNCHROMODE is a Horizon Europe project that aims to develop a data-driven ICT toolbox for improving the management of transport operations from a multimodal perspective. This is, in essence, a multimodal decision support system that will take the form of interconnected pieces of software.
Developing complex systems requires careful planning and management. Their "life cycle" consists of several phases, starting with the critical stage of requirement elicitation, which establishes the system's foundational needs and expectations.
This paper focuses on the requirement elicitation phase of core interconnected services that are to be offered through the SYNCHROMODE toolbox. Elicitation is achieved both through traditional methods and specifically through the use of use cases as well as through the use of gamification.
During the design phase of the toolbox and as a result of the use cases a set of predefined requirements were identified. Moreover, a requirement elicitation serious game was designed and developed. Subsequently, the game was presented in person during a physical workshop, which took place in Thessaloniki in October 2023, to a diverse range of stakeholders, including experts in traffic management, public transport authorities, MaaS providers, first responders, ITS technology providers, and local authorities.
Both requirement elicitation processes are summarized in the context of this paper. Finally, the requirements generated through traditional means and through gamification are then analyzed and compared.

# 1 Introduction

SYNCHROMODE is a Horizon Europe project that aims to develop a data-driven ICT toolbox for improving the management of transport operations from a multimodal perspective. This is, in essence, a complex system that will take the form of interconnected pieces of software.

Typically, developing such complex systems requires careful planning and management. Their “life cycle” consists of several phases, starting with the critical stage of requirements elicitation, which establishes the system’s foundational needs and expectations.

Requirement elicitation for the SYNCHROMODE Toolbox has been initially explored through more traditional means. Specifically, three case studies are explored within the framework of the SYNCHROMODE project, specifically, one in Madrid (Spain), one in Thessaloniki (Greece), and one in South Holland (the Netherlands). For each of the case studies, use cases and user stories have been developed in order to ascertain system requirements.

Particularly in the case of Thessaloniki, a different approach has been followed in parallel. Taking into account that, as it has been argued, the use of games may operate as a supporting tool in fostering stakeholder participation, commitment, communication, and motivation during the requirements elicitation phase of a project’s development [1]. For that reason, game elements have been inserted into the stakeholder engagement and requirement elicitation phases.

This paper explores the effectiveness of gamified requirement elicitation through its application in the design phase of the SYNCHROMODE Toolbox.

# 2 Methodology

## 2.1 Overview

As already mentioned, traditional use cases have been developed and, to an extent, defined prior to the insertion of gamified elements. The introduction of game elements was undertaken during project workshops that engaged various local stakeholders in order to gain their input on system requirements.

In particular, two workshops were conducted under the umbrella of the Thessaloniki case study requirement elicitation phase of the SYNCHROMODE project. The first Thessaloniki Case Study Workshop took place on-site on the 19th of October 2023. It involved 36 representatives of 17 organizations. The second workshop was conducted online on the 25th of January 2024 and involved 34 representatives of 14 organizations. Several of the representatives attended both workshops, while an effort was made for the sample of represented organizations to remain constant throughout. In both cases, the represented organizations fell under one of the following categories:

- (Public) Transport service providers (e.g., bus service provider, taxi fleet operator)
- Academia/R&D (e.g., university, research center)
- Emergency services (e.g., fire brigade, traffic police, ambulance services)

- Public sector/administration/authority (e.g., municipality, regional government, transport oversight authority)
- Road Operator (e.g., highway infrastructure manager)
- Technology providers (e.g., C-ITS service providers)
- Consultancies
- Other

During the first workshop, participants were involved into two linked interactive sessions. For the first interactive session, a serious game was designed and conducted, as will be described in a later section. This effectively introduced game elements into the typical stakeholder engagement process. Players were split into two groups, and members of each group received a booklet outlining the game's scope, rules, and questions. The grouping remained constant for the second interactive session, which took the form of a cooperative SWOT analysis.

The second workshop, which was conducted online, took place over the Microsoft TEAMS platform, and utilized the Mentimeter platform. For the purpose of the workshop, the attendees were presented with three types of questions. The answers to the first type of questions were provided on a 5-point scale. The answers to the second type of questions were provided on a ranking of pre-identified/defined elements, while the remaining questions were open-ended. This more traditional format of engaging stakeholders was also used, besides for receiving the necessary feedback regarding the provided questions, as a tool for assessing the effectiveness of the methodology utilized during the first workshop.

Furthermore, the results of the gamified workshop were compared with a list of preidentified requirements.

### 2.2 Designing the Gamified Workshop

At its most basic form, game design elements may be grouped into the classic points, badges, and leaderboards triad [2]. Points take the form of measurable evidence of progress within the game, badges are, in essence, a visual representation of this progress, and leaderboards allow for the comparison of players against one another [3]. However, more game elements exist that can prove relevant to the gamification of requirement elicitation. Moreover, each of these elements, as shown in Table 1, interacts in specific ways with players in an effort to promote motivation through various motivation mechanisms (Table 2) and, thus, effective participation.

**Table 1.** Summary of game elements and associated motivation mechanisms (based on: [1][3][4]).

| | Game element | Description | Associated motivation drivers |
|---|---|---|---|
| 1 | Activity feed | A stream of recent actions of the player base | Power, Order, Status |
| 2 | Animation | Animated characters used to introduce personas | Curiosity, Order, Tranquility |
| 3 | Avatar | Graphical representation of the current player | Power, Independence, Status |
| 4 | Badges | Visualization of achievements | Power, Order, Saving |
| 5 | Challenges | Steps towards a goal, which are rewarded with badges and points | Curiosity, Independence, Power |
| 6 | Endorsements | Public or social recognition of users' skills or achievements | Acceptance |
| 7 | Exploration | Encourages users to discover new features or content within the game or system | Curiosity |
| 8 | Game master | A moderator of the game | Curiosity, Social Contact, Status |
| 9 | Group forming | The ability to create or join teams | Social Contact |
| 10 | Leaderboard | A ranking of players | Power, Order, Status |
| 11 | Levels | Phases of difficulty in a game to enable progression | Order, Independence Status |
| 12 | Liking | A feature to allow users to support certain content and activities | Power, Status, Vengeance |
| 13 | Onboarding | Process of getting familiar with the game | Curiosity, Independence, Tranquility |
| 14 | Points | the basic means of rewarding users for their activities | Order, Status, Saving |
| 15 | Prize | Physical award given to the winner of the game | Power, Independence, Status |
| 16 | Progress bar | A bar showing the player's current stage in the process | Order, Tranquility |
| 17 | Quiz | A test to let players check their newly acquired knowledge | Curiosity, Independence, Order |
| 18 | Ranking | Related to leaderboard but focuses more on individual status relative to others, rather than a collective display | Power, Status |
| 19 | Resources | Elements that users can collect and use within the game context | Saving |
| 20 | Reward | Covers a broader range of incentives than just prizes, which are typically physical or tangible rewards | Saving, Status |
| 21 | Roles | Defines specific functions or characters that users can adopt | Independence, Power |

| 22 | Scores | Similar to points but often represent a cumulative measure of achievement | Achievement |
|---|---|---|---|
| 23 | Storytelling | A background narrative to arouse positive emotions | Curiosity, Independence, Tranquility |
| 24 | Timer | A countdown showing the remaining time and introducing pressure | Order, Tranquility |
| 25 | Video | Media used to explain user stories and cases | Curiosity, Order, Tranquility |

**Table 2.** Description of the 16 drivers of motivation (based on: [4]).

| | Motivation mechanism | Description |
|---|---|---|
| 1 | Acceptance | The need to be approved and appreciated by others. |
| 2 | Curiosity | The drive to learn and gain knowledge. |
| 3 | Eating | The importance placed on food and enjoyment of eating. |
| 4 | Family | The desire to raise one's own children and spend time with family. |
| 5 | Honor | A commitment to a moral code, often influenced by one's community or religion. |
| 6 | Idealism | The desire for social justice and fairness. |
| 7 | Independence | The need for autonomy and self-reliance. |
| 8 | Order | The preference for organized, predictable, and controlled environments. |
| 9 | Physical Activity | The need for exercise and physical movement |
| 10 | Power | The desire to influence others and seek positions of leadership. |
| 11 | Romance | The desire for sex and beauty. |
| 12 | Saving | The urge to collect items that one perceives as valuable. |
| 13 | Social Contact | The need for friends and companionship. |
| 14 | Status | The pursuit of social standing and importance. |
| 15 | Tranquility: | The desire for emotional calmness and safety. |
| 16 | Vengeance | The drive to compete, win, and retaliate. |

However, for the purposes of requirement elicitation, there is no one-size-fits-all solution because [5]:

- Game elements add an extra layer of technical and organizational costs and expenses.
- The design and use of games requires expertise from different domains, such as behavioral economics, psychology, and human-computer interaction.

- Engagement through games may have an adverse effect on the commitment of some types of stakeholders (introverts, performance, participation solely to win, etc.).
- Any adopted game elements should be adapted to the characteristics and preferences of different stakeholders' characteristics, such as age, gender, culture, and competitiveness.

As indicated in Fig. 1, not only is the choice of the appropriate elements of the aforementioned game important, but furthermore, the interaction between the player, user interface, gameplay, mechanics, and boundaries of the game are of significant importance. An equilibrium needs to be achieved between game complexity and required skill. In this way, both the flow of the game as well as its efficiency in achieving its design purpose is more likely. Without a proper balance, the use of game elements may result, among others, in anxiety, boredom, maladaptation, and, finally, limited or misleading results.

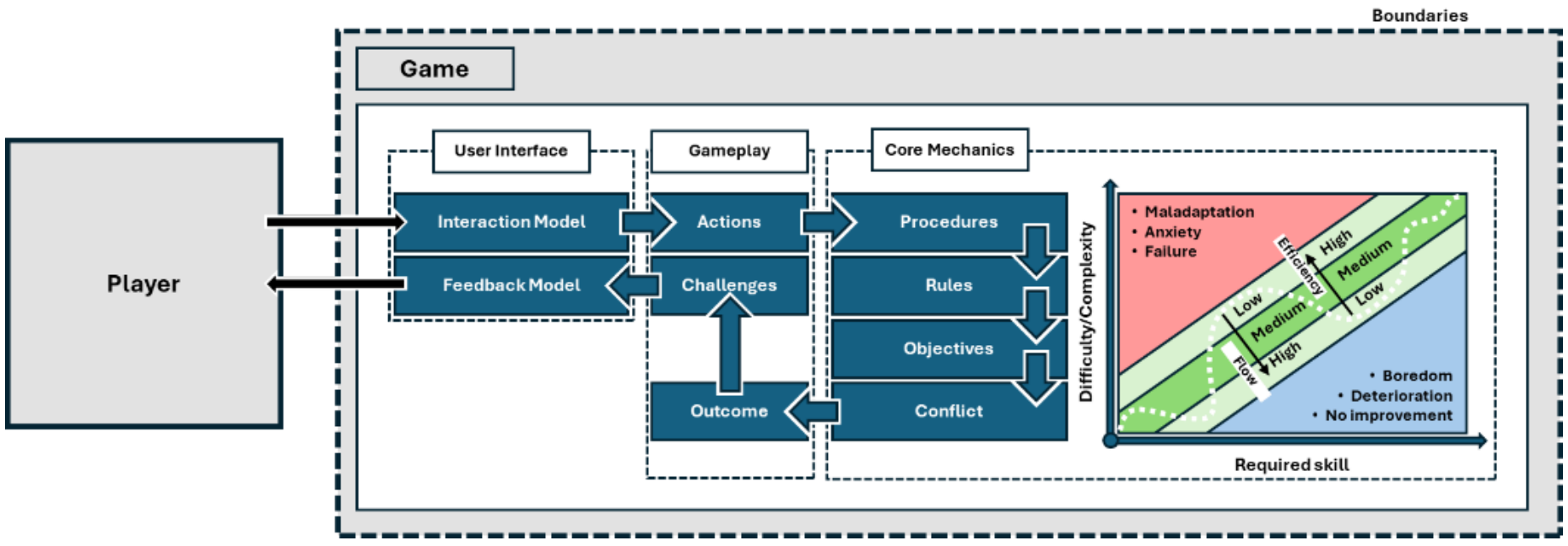

**Fig. 1.** Interactions between game elements and mechanics and effects of game complexity (based on [6] [7] [8]).

Taking into account the nature of the invited stakeholders, it was deemed appropriate for the implemented game to be slightly simpler and to encompass fewer and more to the-point-game elements. Specifically, the final version of the game was as follows:

- Participants were divided into two Groups.
- Each Group was comprised of varying Stakeholders and two Game Masters.
- The members of Group were given a Player Booklet and shown a board depicting the scenario to be played.
- The booklets contained analytical descriptions of the scenarios as well as the rules of the game.
- Questions were asked in sequence to the participants.
- The questions were either yes or no, five-point scales, or open.
- On open questions players were graded.
- Identifying an existing requirement granted them 1 point, while discovering a new one 3 points.
- On these questions the players also voted on the best answers and granted the corresponding player 1 point.

- Scoring was kept and communicated to the players during the game.
- Winning players were announced at the end of the session.

Based on the aforementioned, the gamified workshop incorporated the following game elements as indicated in Table 3.

**Table 3.** Game elements utilized in the gamified workshop.

| | Game element | Description of implementation |
|---|---|---|
| 1 | Challenges | The participants were presented with separate questions and had to engage with each. |
| 2 | Endorsements | The participants voted on the most complete and/or innovating answer to each of the graded questions. |
| 3 | Points | Points were attributed to participants based on whether they provided answers that included preidentified or valid new requirements |
| 4 | Leaderboard | After each graded question the top ranked players were named. |
| 5 | Onboarding | The players were provided with a booklet describing the game and guided through it |
| 6 | Prize | The players were promised a physical prize for the top ranked player |
| 7 | Progress bar | The questions were numbered and the players new the total number of questions |
| 8 | Roles | Each participant was asked to answer based on the role of the stakeholder they represented |
| 9 | Reward | Praise was given to players for innovative answers or group engagement |
| 10 | Storytelling | The players were provided with a map and a detailed story |
| 11 | Timer | A limited amount of time was provided for the answering of each question. The time was communicated to the players. |
| 12 | Game master | Two moderator for each of the parallel games were present. |

# 3 Conducting the Gamified Workshop

As already mentioned, the gamified workshop took place at a physical location with all participants on site. Participants were guided through the process of the game (Fig.2) as described in the methodology section. Participants, after being separated into two groups, were presented with two different scenarios:

- one group addressed an accident on the Thessaloniki Ring-Road during morning peak hours, while
- the other dealt with traffic congestion at the city’s West entrance during evening peak hours.

The gamified session was followed by a cooperative SWOT Analysis, for which the groups remained the same.

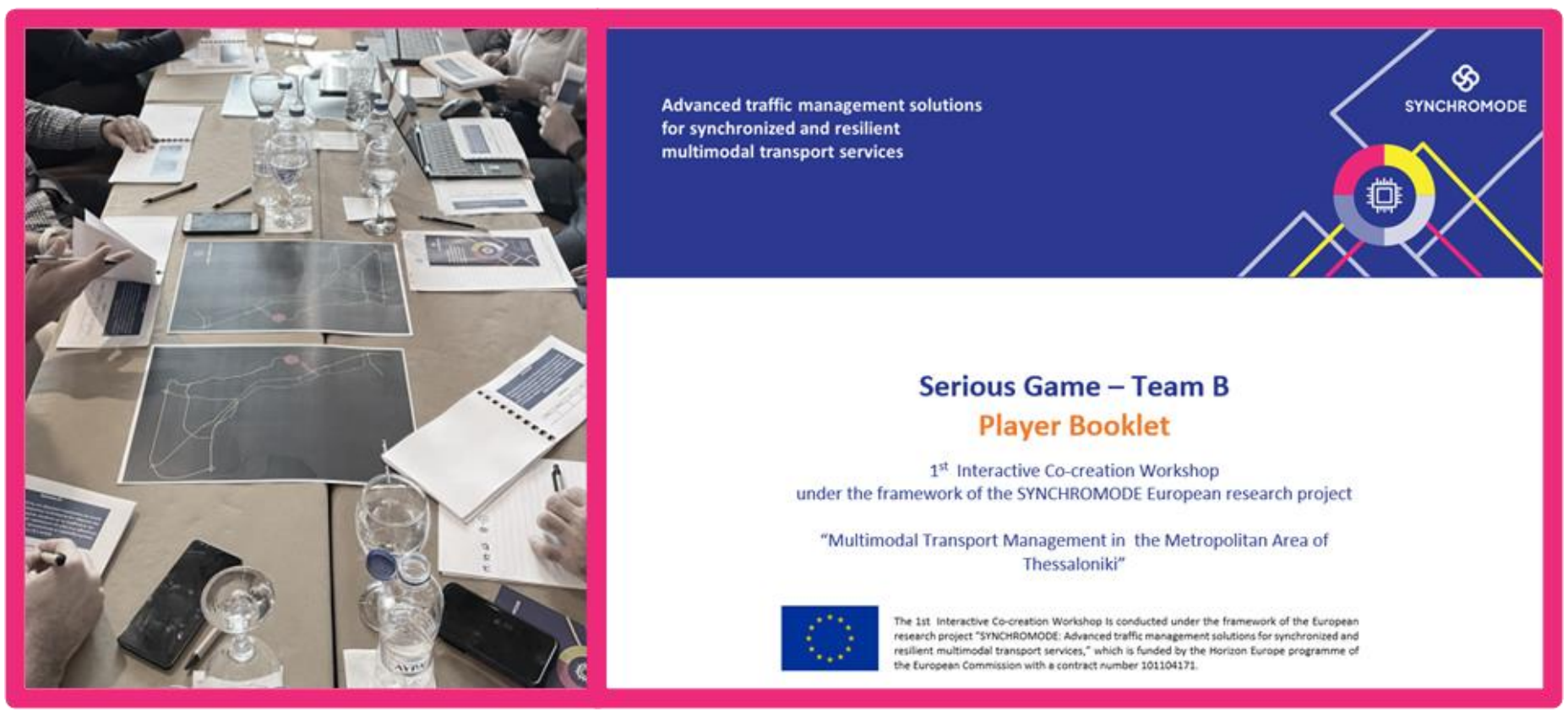

**Fig. 2.** The participants being presented with the booklet and actively participating in the gamified session.

In total, the themes tackled during the gamified session participants were with:

- Resilient traffic management - event management
- Combined passenger and freight transport management
- Traffic management and MaaS
- Coordination between multiple TMCs and/or TMCs and service providers
- Multimodal load balancing
- Real time data exchange between various actors

## 4 Results and Evaluation of the Gamified Workshop

Before the undertaking of the gamified workshop a list of 24 requirements were preidentified through traditional requirement elicitation means, such as user stories and use cases.

While these requirements were incorporated into the game mechanics as a scoring method for providing players with points, it is worth noting that most pre-identified requirements (22 out of 24) were validated through the responses to the scenario-based questions. This high validation rate suggests that the initial assumptions about stakeholder needs were largely accurate.

Moreover, a number of new requirements are deemed as appropriate to introduce to subsequent design stages. Several of these new requirements were also taking into account by the participants when voting on the best answer, further emphasizing their recognition as valid.

For each of the closed-type questions and for each category of stakeholders a consensus index was calculated. This was based on methodology suggested by Tastle and Wierman [9]. The consensus index ranges from 0 to 1 with higher scores indicating more consensus. As such, in the examples included in Fig. 3, a consensus index of 1 indicates unanimity.

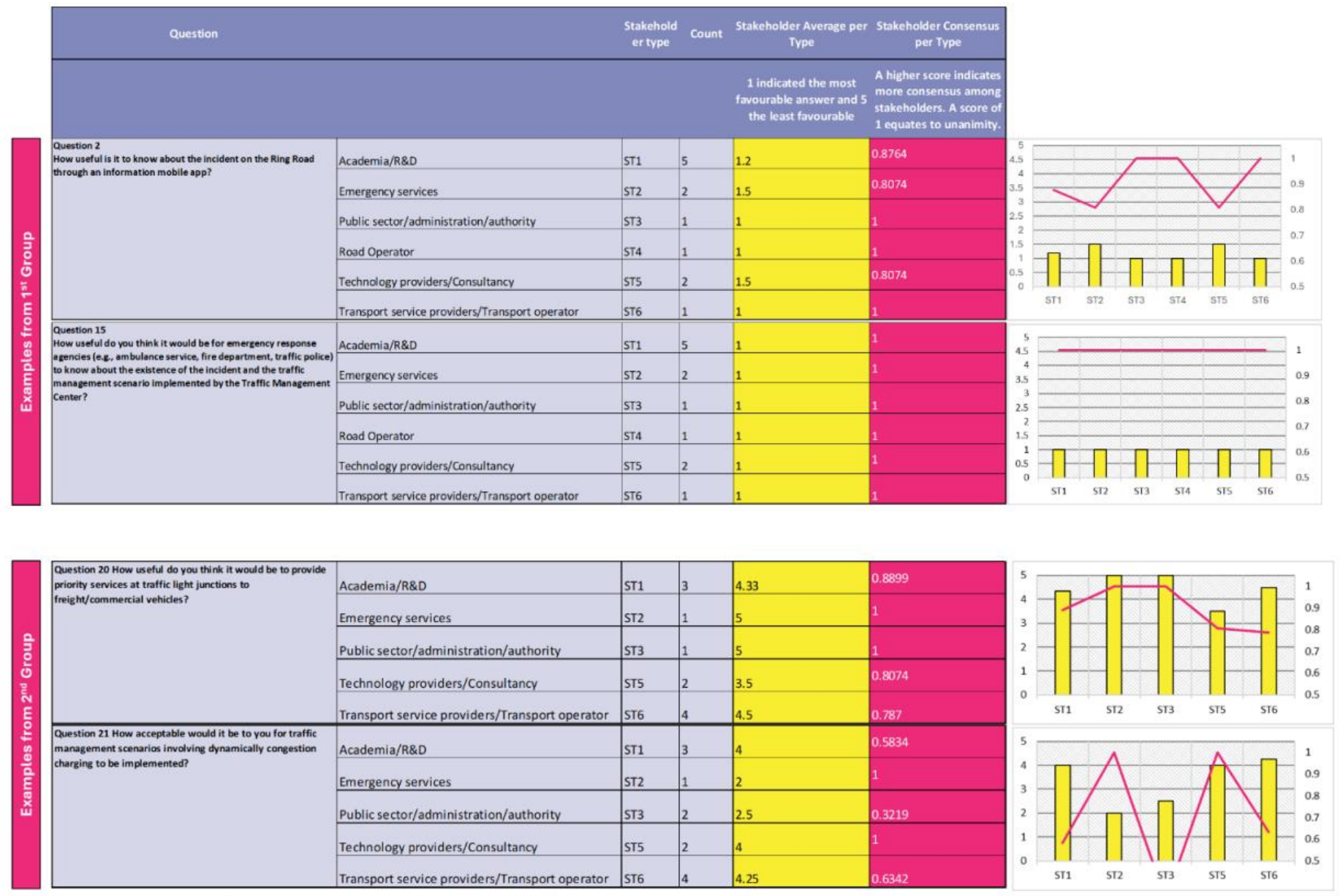

|  | Question | | Stakeholder type | Count | Stakeholder Average per Type | Stakeholder Consensus per Type |
|---|---|---|---|---|---|---|
| | | | | | 1 indicated the most favourable answer and 5 the least favourable | A higher score indicates more consensus among stakeholders. A score of 1 equates to unanimity. |
| Examples from 1st Group | Question 2 How useful is it to know about the incident on the Ring Road through an information mobile app? | Academia/R&D | ST1 | 5 | 1.2 | 0.8764 |
| | | Emergency services | ST2 | 2 | 1.5 | 0.8074 |
| | | Public sector/administration/authority | ST3 | 1 | 1 | 1 |
| | | Road Operator | ST4 | 1 | 1 | 1 |
| | | Technology providers/Consultancy | ST5 | 2 | 1.5 | 0.8074 |
| | | Transport service providers/Transport operator | ST6 | 1 | 1 | 1 |
| | Question 15 How useful do you think it would be for emergency response agencies (e.g., ambulance service, fire department, traffic police) to know about the existence of the incident and the traffic management scenario implemented by the Traffic Management Center? | Academia/R&D | ST1 | 5 | 1 | 1 |
| | | Emergency services | ST2 | 2 | 1 | 1 |
| | | Public sector/administration/authority | ST3 | 1 | 1 | 1 |
| | | Road Operator | ST4 | 1 | 1 | 1 |
| | | Technology providers/Consultancy | ST5 | 2 | 1 | 1 |
| | | Transport service providers/Transport operator | ST6 | 1 | 1 | 1 |
| Examples from 2nd Group | Question 20 How useful do you think it would be to provide priority services at traffic light junctions to freight/commercial vehicles? | Academia/R&D | ST1 | 3 | 4.33 | 0.8899 |
| | | Emergency services | ST2 | 1 | 5 | 1 |
| | | Public sector/administration/authority | ST3 | 1 | 5 | 1 |
| | | Technology providers/Consultancy | ST5 | 2 | 3.5 | 0.8074 |
| | | Transport service providers/Transport operator | ST6 | 4 | 4.5 | 0.787 |
| | Question 21 How acceptable would it be to you for traffic management scenarios involving dynamically congestion charging to be implemented? | Academia/R&D | ST1 | 3 | 4 | 0.5834 |
| | | Emergency services | ST2 | 1 | 2 | 1 |
| | | Public sector/administration/authority | ST3 | 2 | 2.5 | 0.3219 |
| | | Technology providers/Consultancy | ST5 | 2 | 4 | 1 |
| | | Transport service providers/Transport operator | ST6 | 4 | 4.25 | 0.6342 |

**Fig. 3.** Examples of analysis of closed gamified session questions.

The consensus among participants during the workshop was notably high, indicating a strong agreement on several critical aspects of traffic management and incident response. As for particular questions the calculated consensus was higher than that of the second workshop, it is believed that this consensus was to an extent facilitated by the gamified format, which encouraged more active participation and dialogue.

Upon completion of the workshop, participants were sent an interactive questionnaire. Fig. 4 to 7 show the result of the workshop evaluation questions.

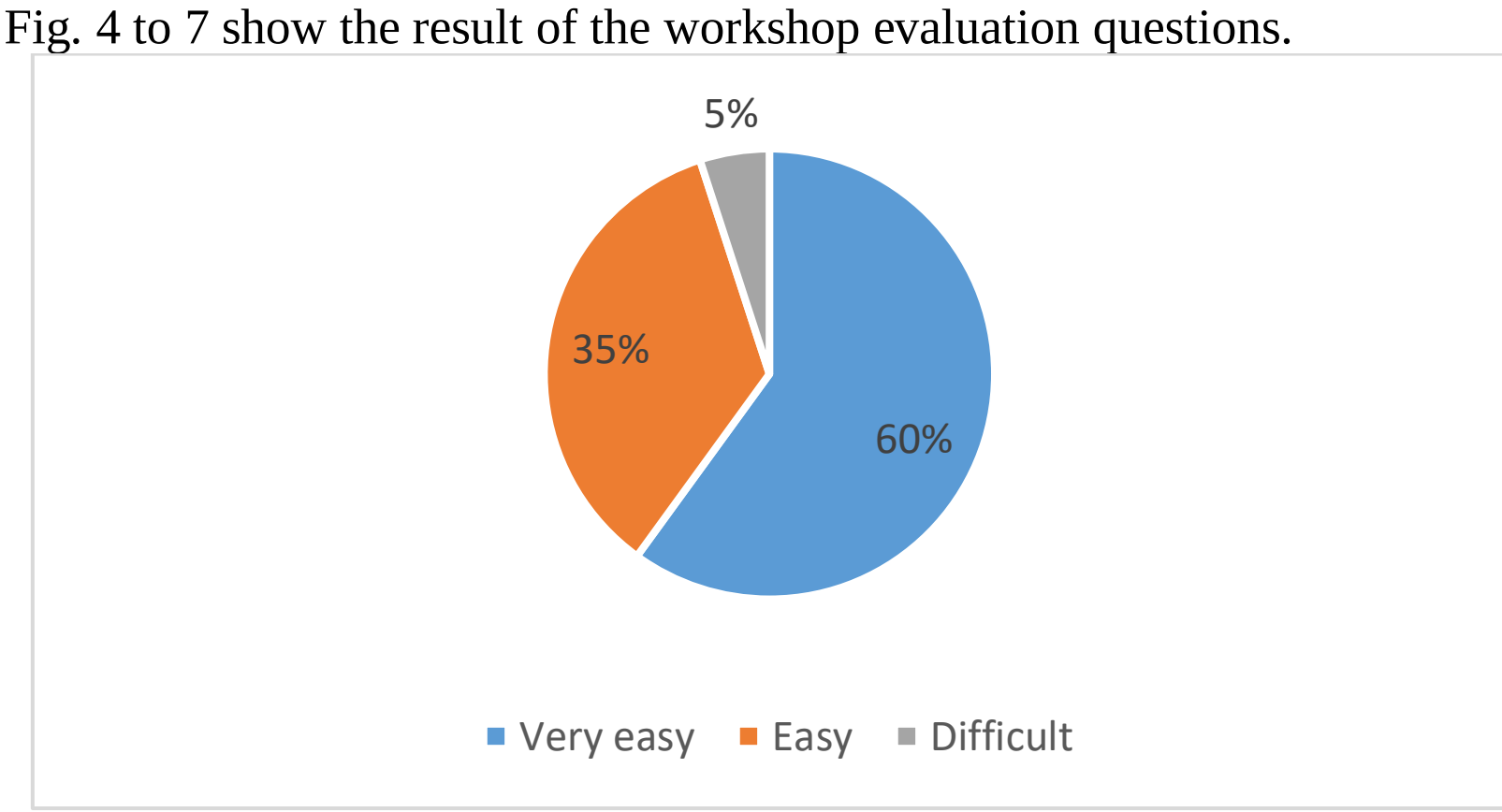

**Fig. 4.** Responses on whether the game rules were easy to follow.

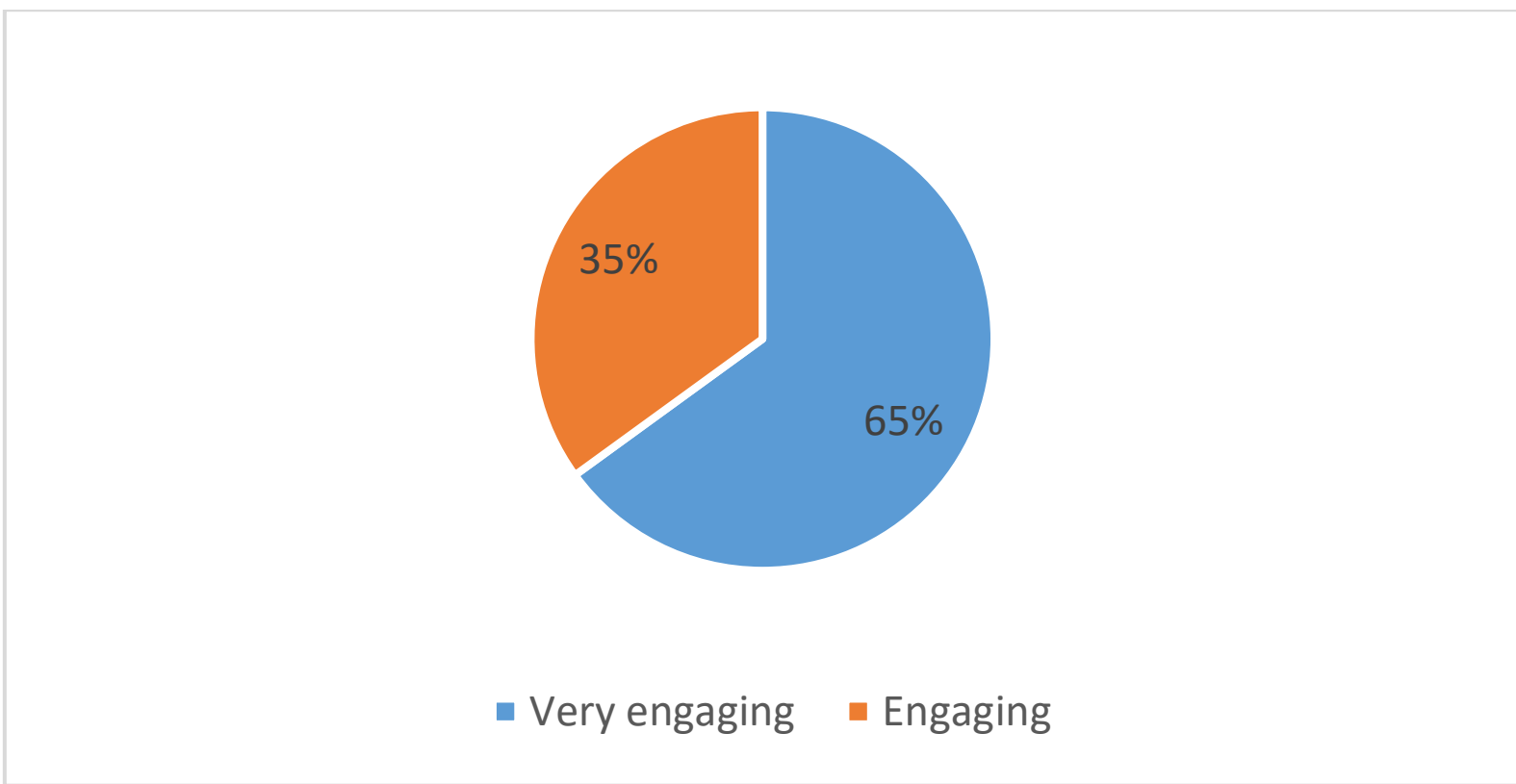

**Fig. 5.** Responses on whether the game was engaging.

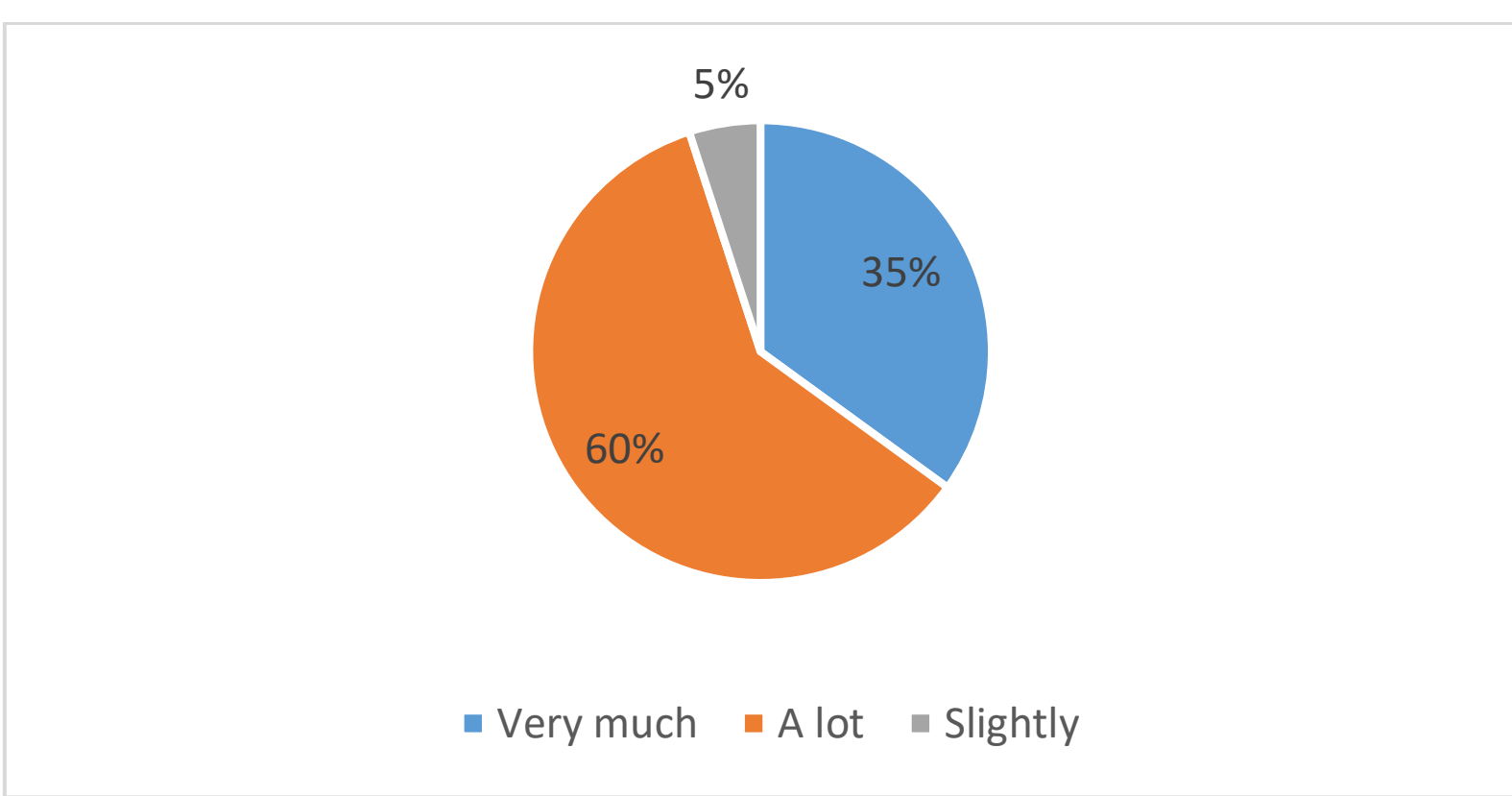

**Fig. 6.** Responses on whether the game promoted the respondent's participation towards identifying requirements and needs.

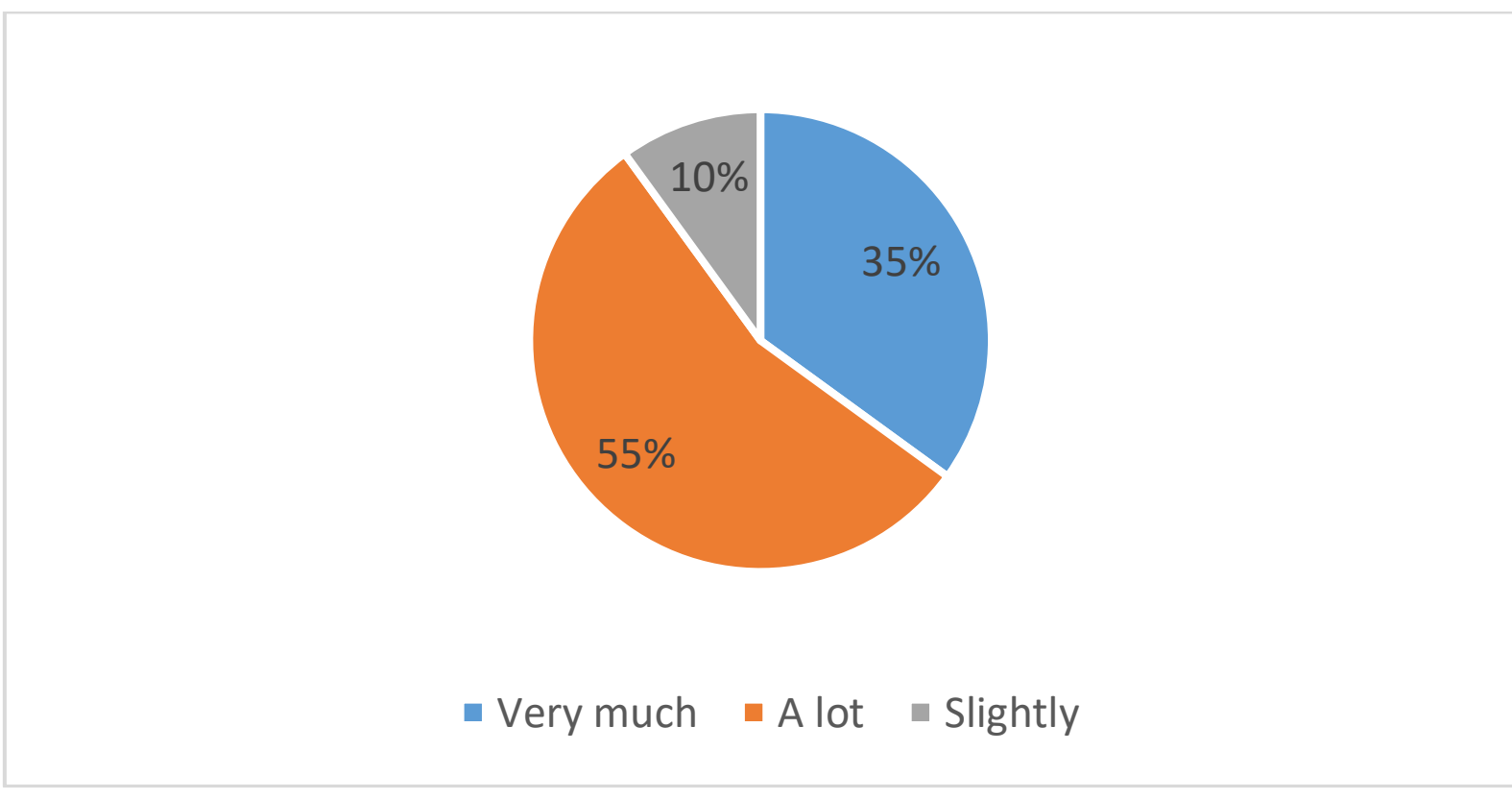

**Fig. 7.** Responses on whether the game effectively promoted the identification of requirements and needs.

Finally, participants of both workshops were asked during the evaluation questionnaire of the second workshop which type of co-creation activity the believed to be more effective. The vast majority of respondents (79%) stated that the gamified workshop was more effective in the co-creation process of requirement elicitation.

## 5 Discussion

This section provides a final reflection on the findings of the gamified requirements elicitation approach and techniques applied during the Greek Case study of the SYNCHROMODE project. The results produced clearly indicate that integrating game elements into the requirement elicitation process not only enhances stakeholder engagement but also facilitates a deeper understanding of stakeholder needs than traditional requirement elicitation methods.

The high validation rate of pre-identified requirements and the identification of new requirements that were often voted as of particular importance during the gamified workshops signify the effectiveness of this approach in capturing comprehensive stakeholder inputs.

Moreover, the consensus indices calculated for responses obtained through gamified interactions indicate that there is often a strong alignment among stakeholders. This suggests that the game elements of the methodology may have helped reduce barriers to expression and encouraged more open communication. This aligns with prior research that has suggested that gamification can increase motivation and participation levels among diverse participant groups.

However, challenges remain. In particular, there is the ever constant need to balance the complexity of game design with the ease of participation to ensure inclusivity and optimal results. Our findings suggest that future applications should ideally include tailor gamification elements tailored to the specific characteristics of the invited stakeholder group, considering factors such as age, cultural background, and professional expertise to maximize engagement and effectiveness.

In conclusion, the study seems to reinforce the potential of gamification as a powerful tool in the arsenal of requirements engineers. This appears particularly true for complex, multi-stakeholder systems like SYNCHROMODE. Further research should explore the scalability of this approach and its applicability across different cultural contexts, as, in this case, it was utilized in only one of the available geographical regions. This could help pave the way for its broader adoption in both academic and industrial settings.

## 6 Acknowledgements

This project has received funding from the European Union’s Horizon Europe research and innovation program under grant agreement No 101104171.